\documentclass[conference, a4paper]{IEEEtran}
\IEEEoverridecommandlockouts
\pdfoutput=1
\usepackage{multirow}
\usepackage{booktabs}
\usepackage{cite}
\usepackage{amsmath,amssymb,amsfonts}
\usepackage{algorithmic}
\usepackage{textcomp}
\usepackage{xcolor}
\usepackage{color}
\ifCLASSINFOpdf
   \usepackage[pdftex]{graphicx}
  \graphicspath{{../pdf/}{../jpeg/}}
   \DeclareGraphicsExtensions{.pdf,.jpeg,.png}
\else
  \usepackage[dvips]{graphicx}
   \graphicspath{{../eps/}}
   \DeclareGraphicsExtensions{.eps}
\fi
\usepackage{array}
\usepackage{epstopdf}
\usepackage{wrapfig}
\usepackage{subfigure}
\usepackage{algorithmic,algorithm}
\def\BibTeX{{\rm B\kern-.05em{\sc i\kern-.025em b}\kern-.08em
    T\kern-.1667em\lower.7ex\hbox{E}\kern-.125emX}}
\begin{document}

\title{Towards THz-based Obstacle Sensing: A Generative Radio Environment Awareness Framework\\
\thanks{This work was supported in part by the National Natural Science Foundation of China under Grant U21B2014 and under Grant 62271121. (Corresponding author: Lingxiang Li)}
}

\author{\IEEEauthorblockN{Tianyu Hu, Yunhang Xie, Shuai Wang, Boyu Ning, Lingxiang Li, Zhi Chen}
\IEEEauthorblockA{\textit{National Key Laboratory of Wireless Communications}\\
\textit{University of Electronic Science and Technology of China (UESTC), Chengdu 611731, China}\\
Emails: \{huty, xieyh\}@std.uestc.edu.cn; shuaiwang@uestc.edu.cn;\\
boydning@outlook.com, lingxiang{\_}li{\_}uestc@hotmail.com; chenzhi@uestc.edu.cn}
}

\maketitle

\begin{abstract}

Obstacle sensing is essential for terahertz (THz) communication since the subsequent beam management can avoid THz signals blocked by the obstacles. In parallel, radio environment, which can be manifested by channel knowledge such as the distribution of received signal strength (RSS), reveals signal propagation situation and the corresponding obstacle information. However, the awareness of the radio environment for obstacle sensing is challenging in practice, as the sparsely deployed THz sensors can acquire only little a priori knowledge with their RSS measurements. Therefore, we formulate in this paper a radio environment awareness problem, which for the first time considers a probability distribution of obstacle attributes. To solve such a problem, we propose a THz-based generative radio environment awareness framework, in which obstacle information is obtained directly from the aware radio environment. We also propose a novel generative model based on conditional generative adversarial network (CGAN), where U-net and the objective function of the problem are introduced to enable accurate awareness of RSS distribution. Simulation results show that the proposed framework can improve the awareness of the radio environment, and thus achieve superior sensing performance in terms of average precision regarding obstacles' shape and location.


\end{abstract}


\section{Introduction}

Recent years have witnessed a notable trend of terahertz (THz) communication standing out as a pivotal technology for future sixth-generation (6G) wireless networks \cite{ning2023beamforming}. Nevertheless, the intrinsic characteristics of THz signals, marked by narrow beams and weak diffraction performance, expose them to frequent obstruction by obstacles\cite{chen2021coverage}, thereby jeopardizing the coverage and quality of THz communication signals. Consequently, the demand for obstacle sensing becomes imperative, ensuring that THz communication systems effectively prevent the resulting disruptions in communication links by employing strategic interventions such as beam switching, relaying, or hand-off \cite{pang2023cellular,charan2021vision,9771340}.





\begin{figure}[!t]
\centering
\includegraphics[width=7cm,height=3cm] {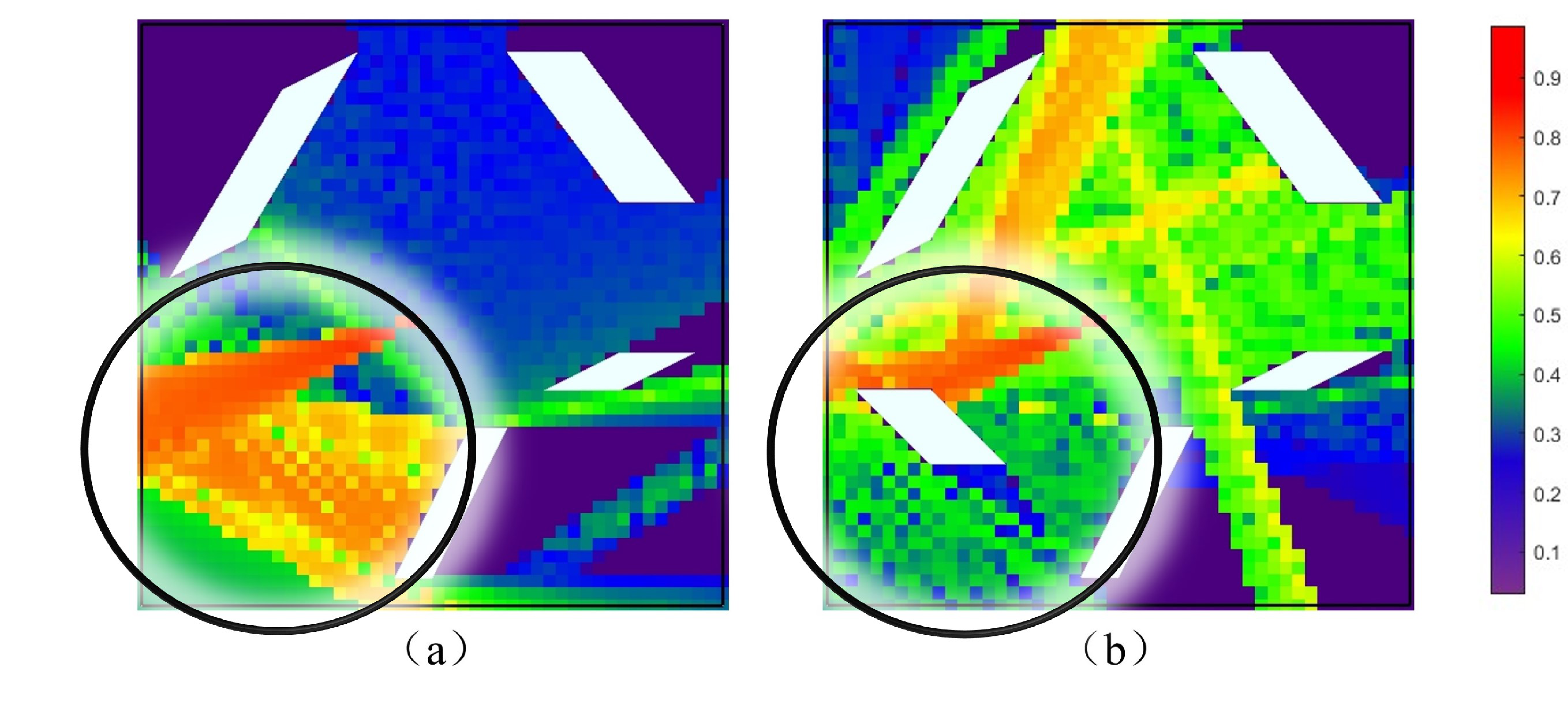}
\caption{THz-based radio environments with the same beam direction, where each block represents an obstacle and different colors indicate the strength of RSS. (a) Without the lower-left obstacle. (b) With the lower-left obstacle.}
\label{area_example}
\end{figure}

Radio environment offers valuable insight into the propagation situation of wireless signals affected by physical environment, particularly obstacles \cite{9540909,9146281,9954187,9140329}. Typically, the radio environment can be manifested through certain representative channel knowledge corresponding to all transceiver locations, such as the distribution of received signal strength (RSS) \cite{10001045}. By intuition, as the strength of RSS affected by obstacles provides direct access to the shape and location information of obstacles, the radio environment characterized by RSS distribution can be used for obstacle sensing.

Given the high directionality of THz beams, the radio environment for THz band may be particularly suitable for obstacle sensing. The reason lies in that the strength of RSS measured in such a band can be strongly related to the existence of obstacles on the line-of-sight (LOS) THz link. Some evidence can be seen in Fig. \ref{area_example}, which illustrates two THz-based radio environments represented by RSS distributions. From the areas boxed by the black line in Fig. \ref{area_example}, one can observe that the obstacle in the lower left corner of Fig. \ref{area_example}(b) does not appear in Fig. \ref{area_example}(a), obviously uncovering the dramatic suppression imposed by such obstacle on the THz beam and the corresponding RSS.

However, in practice, it is challenging to achieve the complete acquisition of the RSS distribution, i.e. the awareness of THz-based radio environment, for obstacle sensing. This is attributed to the fact that it is impractical to position THz sensors (e.g., user equipment) measuring RSS at every conceivable location \cite{xu2021radio}. Hence, merely a little and quite limited prior knowledge can be obtained from the sparsely and randomly deployed sensors. How to overcome this knowledge limitation and derive a comprehensive propagation situation that enables obstacle sensing remains an open research problem.





Although there have been some related studies \cite{xu2021radio,levie2021radiounet,zeng2022uav,hu20233d}, they cannot achieve the desired awareness of THz-based radio environment considering THz channel characteristics (e.g., molecular absorption) and the impact of obstacles on signal propagation with sparse sensors. Specifically, \cite{xu2021radio} considered a statistical channel propagation model, but the resulting radio environment will be wrong when deterministic obstacles are present. \cite{levie2021radiounet} proposed a radio environment awareness approach named RadioUNet, but it depends on the strong assumption of known obstacles. The method proposed by \cite{zeng2022uav} achieved the awareness of a certain scenario but using densely and evenly deployed sensors. In addition, \cite{hu20233d} proposed DCRGAN to generate the accurate distribution of RSS by exploiting generative adversarial networks (GANs) \cite{goodfellow2014generative}, which however only considered signal propagation in free space without obstacles and the unique THz characteristics, leading to unsatisfactory radio environment awareness.

%



%
In this paper, recognizing the relevance of obstacle information to the THz-based radio environment, and motivated by the ability of generative models, such as GANs, we aim to delve into the necessary yet challenging task of THz-based radio environment awareness based on a generative model for obstacle sensing. Our contributions are summarized as follows:

\begin{itemize}
	\item We formulate a radio environment awareness problem considering a probability distribution of obstacle attributes (e.g., obstacles' shapes, numbers, and locations). To the authors' knowledge, this is the first time the expectation of awareness performance corresponding to such a probability distribution has been considered, instead of pursuing awareness performance for a particular scenario.
	\item We propose a THz-based generative radio environment awareness framework to sense obstacles with arbitrary attributes from the perspective of radio environment segmentation, where the formulated problem is solved by the generative model to obtain the aware radio environment.
	\item We propose a novel generative model based on conditional GAN (CGAN) \cite{isola2017image}. With a priori knowledge as a condition, the proposed model can learn the THz signal propagation mechanism w.r.t. RSS distribution. Moreover, we introduce the objective function of the problem and U-net \cite{ronneberger2015u} to improve the awareness performance of the radio environment and the corresponding segmentation performance, respectively.
	\item Simulation results show that the proposed framework achieves accurate awareness of radio environments in terms of the weighted mean squared error (MSE). Furthermore, for obstacle sensing with concern, the proposed framework outperforms baselines in terms of average precision (AP) \cite{ren2015faster} regarding obstacles' shape and location.
\end{itemize}

\section{System Model}

\begin{figure}[!t]
\centering
\includegraphics[width=3.8cm,height=3.5cm] {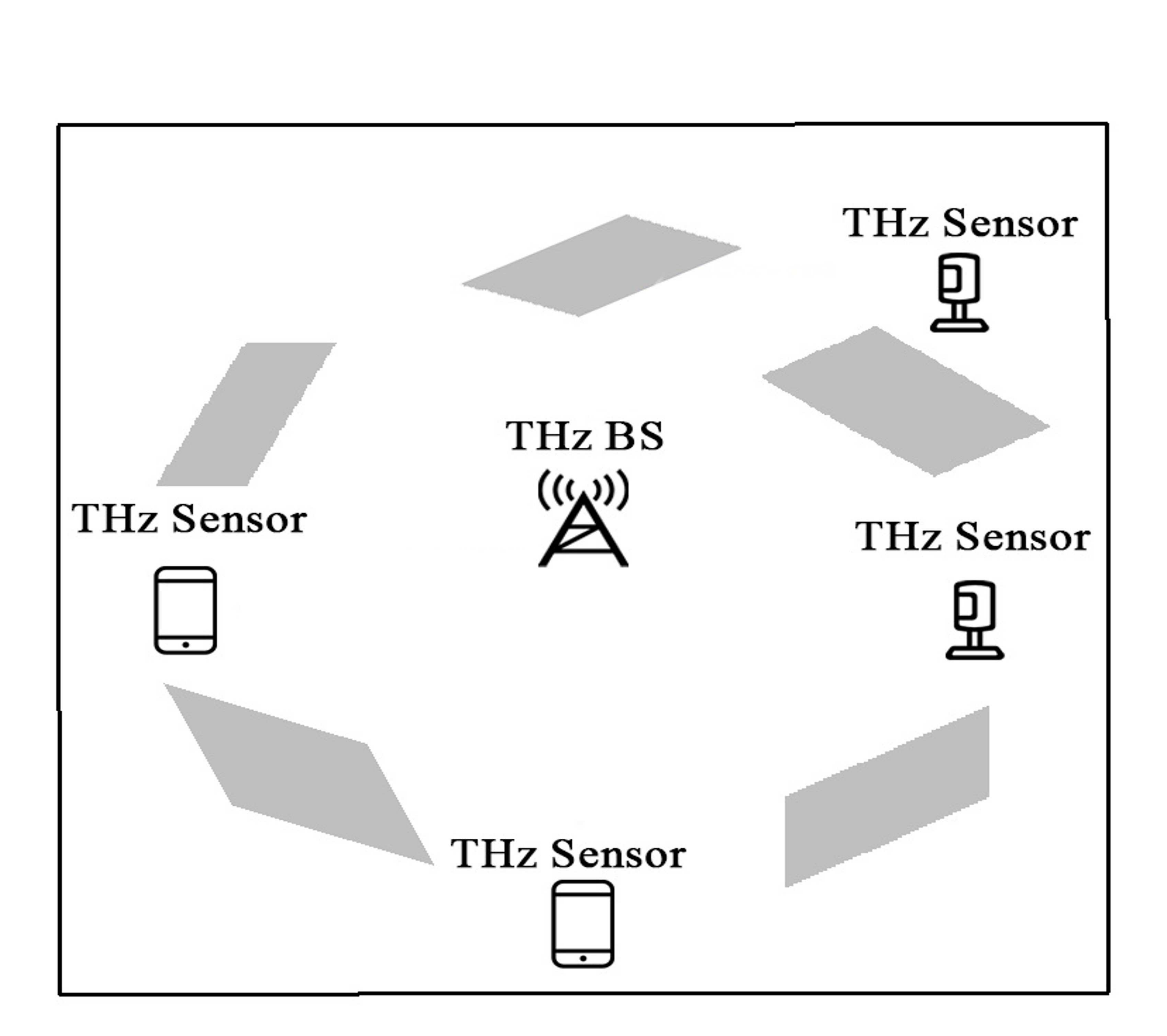}
\caption{An example of the considered scenarios for obstacle sensing.}
\label{new_scenario}
\end{figure}

We consider a number of two-dimensional (2D) scenarios with the same length and width, where each scenario is evenly divided into $N_{\mathrm{A}}=N_{\mathrm{L}}\times N_{\mathrm{W}}$ grids. In each scenario, the obstacles have random quadrilateral shapes and positions, and the number of obstacles is also random. In addition, we assume that the obstacles do not overlap each other. As a result, we can define $3$ random variables $Z_{\mathrm{O_1}}$, $Z_{\mathrm{O_2}}$, and $Z_{\mathrm{O_3}}$ related to obstacle attributes, i.e. the shape, number, and location of the obstacles, respectively. Hence, $\left(Z_{\mathrm{O_1}},Z_{\mathrm{O_2}},Z_{\mathrm{O_3}}\right) \sim P$, where $P$ is the joint probability distribution of these random variables. For notational simplicity, we define a random vector $\mathbf{z}_{\mathrm{O}}=\left[Z_{\mathrm{O_1}},Z_{\mathrm{O_2}},Z_{\mathrm{O_3}}\right]^T$.

Moreover, a THz base station (BS) located at the center $\mathbf{x}_{\mathrm{tx}}$ plays the role of communication transmitter, while $K$ THz sensors are randomly and sparsely distributed with locations $\{\mathbf{x}_{\mathrm{rx},k}\}_{k=1}^{K}$ and communicate with BS as receivers. Hence, the considered RSS for characterizing the radio environment relies on the measurement of the downlink signal at THz sensors, and the corresponding awareness can be achieved by a central processing unit at BS, where the measured RSS as a priori knowledge should be uploaded via uplink signal. To compensate for the severe path loss of the THz channel \cite{ning2023beamforming}, the BS is equipped with uniform circular arrays (UCA), where the generated high-gain directional beam has nearly identical beamwidth in any direction \cite{jiao2019millimeter}, while sensors are equipped with omnidirectional antennas.

\subsection{Signal Propagation Model} 

For a given transmitted THz signal $\mathbf{x}(t)$ and the obstacle attributes $\mathbf{z}_{\mathrm{O}}$, the corresponding received signal $y_{k,\mathbf{z}_{\mathrm{O}}}(t)$ at the $k$\,th sensor can be expressed as
\begin{equation}
\begin{aligned}
y_{k,\mathbf{z}_{\mathrm{O}}}(t)=\mathbf{h}_{k,\mathbf{z}_{\mathrm{O}}}(t)\mathbf{x}(t)+z(t),
\label{receive_signal}
\end{aligned}
\end{equation} where $\mathbf{h}_{k,\mathbf{z}_{\mathrm{O}}}(t)$ denotes the time-domain channel vector between the $k$\,th sensor and the BS, and $z(t)$ is the additive white Gaussian noise (AWGN) with variance $\sigma^{2}$. Note that channel $\mathbf{h}_{k,\mathbf{z}_{\mathrm{O}}}(t)$ is highly associated with THz channel characteristics (e.g. molecular absorption \cite{ning2023beamforming}). In addition, $\mathbf{x}(t)=\mathbf{f} s(t)\exp{\left(j2\pi f_{\mathrm{c}}t\right)}$, where $s(t)$, $\mathbf{f}$, and $f_{\mathrm{c}}$ are the baseband signal, the analog beamformer of the transmitted beam, and the carrier frequency, respectively.

To obtain the RSS of sensors, we utilize a practical signal propagation modeling method, i.e. ray tracing, as the conventional signal propagation model \cite{hu20233d} ignores the effect of obstacles on RSS, leading to model mismatch in practice. Specifically, the ray tracing method first calculates the path of each ray corresponding to obstacles and then coherently combines the electric field of each ray path corresponding to the same cluster \cite{remcom}. Hence, by summing the resulting power of each cluster, the RSS $\Psi_{\mathbf{z}_{\mathrm{O}}}\left(\mathbf{x}_{\mathrm{rx}, k}\right)$ corresponding to the $k$\,th sensor and the obstacle attributes $\mathbf{z}_{\mathrm{O}}$ can be given as \begin{equation}
\begin{aligned}
\Psi_{\mathbf{z}_{\mathrm{O}}}\left(\mathbf{x}_{\mathrm{rx}, k}\right)=\Psi\left(\mathbf{x}_{\mathrm{tx}}\right)\Gamma_{\mathbf{x}_{\mathrm{tx}} \rightarrow \mathbf{x}_{\mathrm{rx},k}}+\sigma^{2},
 \label{Model_RSS}
\end{aligned}
\end{equation} where $\Psi\left(\mathbf{x}_{\mathrm{tx}}\right)$ and $\Gamma_{\mathbf{x}_{\mathrm{tx}} \rightarrow \mathbf{x}}$ are the RSS for the BS location $\mathbf{x}_{\mathrm{tx}}$ (i.e. the total transmit power $P_{\mathrm{T}}$) and a function defined to characterize the above ray tracing process, respectively.

\subsection{Problem Formulation of Radio Environment Awareness}

The radio environment can be characterized by a function mapping from the location $\mathbf{x}$ to the RSS $\Psi_{\mathbf{z}_{\mathrm{O}}}(\mathbf{x})$, i.e. the distribution of RSS. In parallel, the RSS at each grid can be calculated by ray tracing, which assumes that THz sensors are deployed on each grid. Therefore, the complete awareness of THz-based radio environment can be achieved by ray tracing, and the resulting matrix $\boldsymbol{\Psi}_{\mathbf{z}_{\mathrm{O}}}=\lbrace \Psi_{\mathbf{z}_{\mathrm{O}}}(\mathbf{x}) \vert \forall \mathbf{x} \!\in \mathcal{C}_{\mathrm{A}} \rbrace$ representing the radio environment can be obtained, where $\mathcal{C}_{\mathrm{A}}$ is the set of all grids and $\boldsymbol{\Psi}_{\mathbf{z}_{\mathrm{O}}}\in \mathbb{R}^{N_{\mathrm{L}}\times N_{\mathrm{W}}}$. Because of the suppression imposed by obstacles on directional THz signals, which is our motivation for segmenting obstacles from radio environment, we consider $\Psi_{\mathbf{z}_{\mathrm{O}}}(\mathbf{x})=0,\forall \mathbf{x}\in \mathcal{C}_{\mathbf{z}_{\mathrm{O}}}$ for the set $\mathcal{C}_{\mathbf{z}_{\mathrm{O}}}$ of grids whose locations are inside the obstacles.

However, such ray tracing calculation is almost impossible in practice due to the high complexity. It is also not possible to densely place sensors for RSS measurements. Actually, only the sparse RSS measurements $\{\Psi_{\mathbf{z}_{\mathrm{O}}}\left(\mathbf{x}_{\mathrm{rx}, k}\right)\}_{k=1}^{K}$ can be exploited for THz-based radio environment awareness. That is, with the available RSS corresponding to $K$ THz sensors, only a prior matrix $\widetilde{\boldsymbol{\Psi}}_{\mathbf{z}_{\mathrm{O}}}=\lbrace\Psi_{\mathbf{z}_{\mathrm{O}}}(\mathbf{x})=\Psi_{\mathbf{z}_{\mathrm{O}}}\left(\mathbf{x}_{\mathrm{rx}, k}\right) \vert \forall \mathbf{x} \in \mathcal{C}_{\mathrm{B}} \rbrace \bigcup \lbrace\Psi_{\mathbf{z}_{\mathrm{O}}}(\mathbf{x})\!=0\vert \forall \mathbf{x} \!\in\! \mathcal{C}_{\mathrm{A}} \backslash \mathcal{C}_{\mathrm{B}} \rbrace$ representing a subset of the radio environment can be obtained, and $\mathcal{C}_{\mathrm{A}} \backslash \mathcal{C}_{\mathrm{B}}$ is the complement of set $\mathcal{C}_{\mathrm{B}}=\{\mathbf{x}_{\mathrm{rx}, 1},\cdots,\mathbf{x}_{\mathrm{rx}, K}\}$ in $\mathcal{C}_{\mathrm{A}}$.

Therefore, the problem of radio environment awareness lies in how to use the prior knowledge $\widetilde{\boldsymbol{\Psi}}_{\mathbf{z}_{\mathrm{O}}}$ to perform awareness for arbitrary scenarios with the distribution $P$ and to approach the complete THz-based radio environment $\boldsymbol{\Psi}_{\mathbf{z}_{\mathrm{O}}}$. Considering an imperfect awareness with a generative model, we can define a matrix $\widehat{\boldsymbol{\Psi}}_{\mathbf{z}_{\mathrm{O}}}=\lbrace \widehat{\Psi}_{\mathbf{z}_{\mathrm{O}}}(\mathbf{x})=\Psi_{\mathbf{z}_{\mathrm{O}}}\left(\mathbf{x}_{\mathrm{rx}, k}\right) \vert \forall\mathbf{x} \in \mathcal{C}_{\mathrm{B}} \rbrace \bigcup \lbrace\widehat{\Psi}_{\mathbf{z}_{\mathrm{O}}}(\mathbf{x}) \vert \forall \mathbf{x} \in \mathcal{C}_{\mathrm{A}} \backslash \mathcal{C}_{\mathrm{B}} \rbrace$ representing the resulting aware radio environment, where $\widehat{\Psi}_{\mathbf{z}_{\mathrm{O}}}(\mathbf{x})$ for $\mathbf{x} \in \mathcal{C}_{\mathrm{A}} \backslash \mathcal{C}_{\mathrm{B}} $ is the estimated RSS values for grids without THz sensor deployment. Note that if such an aware radio environment can be sufficiently accurate, then we can sense the obstacle information more effectively through the strength of the estimated RSS for arbitrary scenarios.

Based on the above, the THz-based radio environment awareness problem can be formulated as
\begin{equation}
\begin{aligned}
 & \underset{\widehat{\boldsymbol{\Psi}}_{\mathbf{z}_{\mathrm{O}}}}{\mathrm{min}}\,\,\, V_{\mathrm{MSE}}=\mathbb{E}_{\widehat{\boldsymbol{\Psi}}_{\mathbf{z}_{\mathrm{O}}}}[f(\widehat{\boldsymbol{\Psi}}_{\mathbf{z}_{\mathrm{O}}})]\\
 & s.t. \,\,\,\, \widehat{\Psi}_{\mathbf{z}_{\mathrm{O}}}(\mathbf{x})=\Psi_{\mathbf{z}_{\mathrm{O}}}(\mathbf{x}), \forall\mathbf{x} \in \mathcal{C}_{\mathrm{B}},\\
 & \,\,\,\,\,\,\,\,\,\,\,\,\,\, \mathbf{z}_{\mathrm{O}} \sim P,
\label{reconstruction_problem}
\end{aligned}
\end{equation}where the joint probability distribution $P$ of obstacle attributes corresponding to random vector $\mathbf{z}_{\mathrm{O}}$ is considered, and $f(\widehat{\boldsymbol{\Psi}}_{\mathbf{z}_{\mathrm{O}}})$ is a distortion function with the form of a weighted MSE, i.e.
\begin{equation}
\begin{aligned}
 f(\widehat{\boldsymbol{\Psi}}_{\mathbf{z}_{\mathrm{O}}})=\frac{1}{N_{\mathrm{A}}}\sum_{\mathbf{x} \in \mathcal{C}_{\mathrm{A}}}w_{\mathbf{z}_{\mathrm{O}}}^2(\mathbf{x})(\Psi_{\mathbf{z}_{\mathrm{O}}}(\mathbf{x})-\widehat{\Psi}_{\mathbf{z}_{\mathrm{O}}}(\mathbf{x}))^{2}.
\label{MSE_problem}
\end{aligned}
\end{equation} Note that $w_{\mathbf{z}_{\mathrm{O}}}(\mathbf{x})$ is the weight value for each grid to indicate the importance of obstacle areas.



\section{THz-Based Generative Radio Environment Awareness Framework}

This section proposes a THz-based generative radio environment awareness framework for obstacle sensing, where the obstacle information can be obtained by the accurate awareness of the radio environment. However, solving problem (\ref{reconstruction_problem}) for radio environment awareness can be challenging, because characterizing the probability density function of the distribution P is intractable, and only
little a priori knowledge is available for each scenario. In the proposed framework, considering the generative model's ability to learn and represent data distributions, we proposed a CGAN-based generative model to solve problem (\ref{reconstruction_problem}), where a priori knowledge, i.e. the sparse RSS measurements are used as a condition.

\begin{figure}[!t]
\centering
\includegraphics[width=9cm,height=3.2cm] {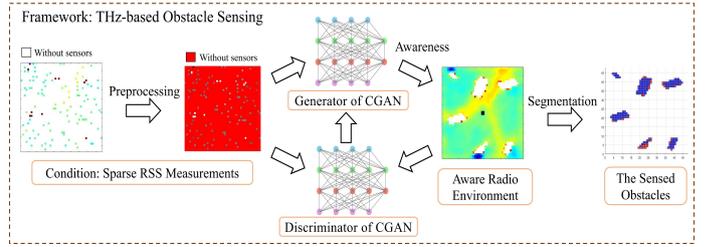}
\caption{The key idea of the proposed framework for THz-based obstacle sensing.}
\label{overall_framework}
\end{figure}

The key idea of the proposed framework is illustrated in Fig. \ref{overall_framework}. Specifically, we first preprocess the sparse RSS measurements obtained from THz sensors. For the proposed generative model, the resulting preprocessing matrix is input to a generator and a discriminator performing adversarial training. Note that the generator does the task of radio environment awareness. After the training, obstacle sensing is performed based on the segmentation of the aware radio environment.


\subsection{Weight-Targeted Preprocessing}

\addtolength{\topmargin}{0.03in}

For obstacle sensing, an important step in solving problem (\ref{reconstruction_problem}) is to design each grid's weight $w_{\mathbf{z}_{\mathrm{O}}}(\mathbf{x})$, by which the solved radio environment $\widehat{\boldsymbol{\Psi}}_{\mathbf{z}_{\mathrm{O}}}$ can reveal the obstacle information in a better way. A straightforward idea is to assign the highest weight to the obstacle areas $\mathcal{C}_{\mathbf{z}_{\mathrm{O}}}$, while the weight of other areas $\mathcal{C}_{\mathrm{A}} \backslash \mathcal{C}_{\mathbf{z}_{\mathrm{O}}} $ is determined by the RSS value of each grid itself, as communication systems tend to focus on hotspot areas with larger RSS when exploiting radio environments.

Specifically, for each grid whose location is inside the obstacles, the corresponding weight is designed to value $1$, i.e. $w_{\mathbf{z}_{\mathrm{O}}}(\mathbf{x})=1,\forall \mathbf{x}\in \mathcal{C}_{\mathbf{z}_{\mathrm{O}}}$. Moreover, for a grid $\mathbf{x}\in \mathcal{C}_{\mathrm{A}} \backslash \mathcal{C}_{\mathrm{O}}$ with the RSS $\Psi_{\mathbf{z}_{\mathrm{O}}}(\mathbf{x})$, the corresponding weight is designed to $w_{\mathbf{z}_{\mathrm{O}}}(\mathbf{x})=\Psi_{\mathrm{smin}}+\Psi_{\mathrm{smax}} \frac{\Psi_{\mathbf{z}_{\mathrm{O}}}(\mathbf{x})-\Psi_{\mathrm{min}}} {\Psi_{\mathrm{max}}-\Psi_{\mathrm{min}}}, \forall \mathbf{x}\in \mathcal{C}_{\mathrm{A}} \backslash \mathcal{C}_{\mathrm{O}}$, where $\Psi_{\mathrm{min}}$ and $\Psi_{\mathrm{max}}$ are the minimum and the maximum values of RSS, respectively, while $\Psi_{\mathrm{smin}}$ and $\Psi_{\mathrm{smax}}$ are values that satisfy $0=\Psi_{\mathrm{smin}}<\Psi_{\mathrm{smax}}<1$. As a result, a weight matrix $\mathbf{W}_{\mathbf{z}_{\mathrm{O}}}$ corresponding to the THz-based radio environment $\boldsymbol{\Psi}_{\mathbf{z}_{\mathrm{O}}}$ can be obtained as a one-channel gray-scale map.

Considering the weight range versus the RSS range, to better solve problem (\ref{reconstruction_problem}) by the generative model, we perform a preprocessing operation similar to the above on the matrix $\boldsymbol{\Psi}_{\mathbf{z}_{\mathrm{O}}}$ and the sparse RSS measurements $\widetilde{\boldsymbol{\Psi}}_{\mathbf{z}_{\mathrm{O}}}$ acting as a priori knowledge. Given that the weight matrix $\mathbf{W}_{\mathbf{z}_{\mathrm{O}}}$ is directly mapped by the complete matrix $\boldsymbol{\Psi}_{\mathbf{z}_{\mathrm{O}}}$, we can treat this weight matrix as the preprocessed complete matrix.

However, for the prior matrix $\widetilde{\boldsymbol{\Psi}}_{\mathbf{z}_{\mathrm{O}}}$, the generative model may have difficulty distinguishing the obstacle areas $ \mathcal{C}_{\mathbf{z}_{\mathrm{O}}}$ from the area $\mathcal{C}_{\mathrm{A}} \backslash \mathcal{C}_{\mathrm{B}}$ without THz sensors deployment, as the RSS corresponding to these areas are all set to $0$. Therefore, we consider that the unknown RSS $\Psi_{\mathbf{z}_{\mathrm{O}}}(\mathbf{x})$ in such non-deployed areas $\mathcal{C}_{\mathrm{A}} \backslash \mathcal{C}_{\mathrm{B}}$ can be mapped to a three-channel vector, such as $\left[1, 0, 0\right]^T$, which shows a red color and can be distinguished from the grayscale used in $\boldsymbol{\Psi}_{\mathbf{z}_{\mathrm{O}}}$ or $\mathbf{W}_{\mathbf{z}_{\mathrm{O}}}$. In addition, the measured RSS $\Psi_{\mathbf{z}_{\mathrm{O}}}(\mathbf{x})$ with $\mathbf{x}\in \mathcal{C}_{\mathrm{B}}$ is linearly mapped to the gray vector $\left[w_{\mathbf{z}_{\mathrm{O}}}(\mathbf{x}), w_{\mathbf{z}_{\mathrm{O}}}(\mathbf{x}), w_{\mathbf{z}_{\mathrm{O}}}(\mathbf{x})\right]^T$. Hence, a priori matrix can be preprocessed as a three-channel but colored map, while the weight matrix and the preprocessed complete matrix also need to be converted to three-channel gray-scale maps. Fig. \ref{overall_framework} demonstrates an example of the sparse RSS measurements (i.e. the prior matrix) before and after preprocessing.



\subsection{CGAN-based Generative Model with Off-Line Training}

For solving the THz-based radio environment awareness problem (\ref{reconstruction_problem}), the difficulties of expressing the distribution $P$ of obstacle attributes and acquiring the RSS data in practice need to be addressed. Therefore, we exploit ray tracing to obtain a number of complete THz-based radio environments offline, which can be considered as an implicit approximation of the RSS distribution under arbitrary obstacle attributes, and thus can mitigate the challenges of practical deployment. In this case, the key to solving problem (\ref{reconstruction_problem}) lies in how to make effective utilization of this large amount of data in a supervised manner, thus ensuring better awareness of the THz-based radio environment and the corresponding obstacle sensing.

Therefore, we propose a CGAN-based generative model in our framework, which exploits the advanced structure of CGAN, i.e. the adversarial training under a prior condition, to enable RSS distribution capturing and radio environment generation under arbitrary obstacle attributes. Specifically, with CGAN, the proposed generative model consists of two sub-networks referred to as a generator and a discriminator, where the prior matrix $\widetilde{\boldsymbol{\Psi}}_{\mathbf{z}_{\mathrm{O}}}$ after preprocessing is used as the condition. The generator can attain a generated matrix $\widehat{\boldsymbol{\Psi}}_{\mathbf{z}_{\mathrm{O}}} =G(\widetilde{\boldsymbol{\Psi}}_{\mathbf{z}_{\mathrm{O}}} ; \mathbf{\Theta}_{g})$ treated as the aware radio environment. While the discriminator \(D(\widehat{\boldsymbol{\Psi}}_{\mathbf{z}_{\mathrm{O}}}\vert \widetilde{\boldsymbol{\Psi}}_{\mathbf{z}_{\mathrm{O}}} \,\mathrm{or}\, \boldsymbol{\Psi}_{\mathbf{z}_{\mathrm{O}}}\vert \widetilde{\boldsymbol{\Psi}}_{\mathbf{z}_{\mathrm{O}}}; \mathbf{\Theta}_{d})\) can obtain a discrimination result about whether the input data (i.e. the generated matrix $\widehat{\boldsymbol{\Psi}}_{\mathbf{z}_{\mathrm{O}}}$ or the complete matrix $\boldsymbol{\Psi}_{\mathbf{z}_{\mathrm{O}}}$) is synthetic data (from the generator) or actual data (from ray tracing and preprocessing) as well as matches the condition $\widetilde{\boldsymbol{\Psi}}_{\mathbf{z}_{\mathrm{O}}}$, where \(\mathbf{\Theta}_{g}\) and \(\mathbf{\Theta}_{d}\) denote the parameters of generator and discriminator, respectively. Furthermore, $G$ aims to generate the samples closer to actual data to fool $D$, and $D$ aims to output a more accurate discrimination result to avoid being fooled. Thus, $D$  and $G$ can enhance each other's performance until the training reaches an equilibrium point.

\begin{figure}[!t]
\center
\includegraphics[width=9cm,height=5.5cm] {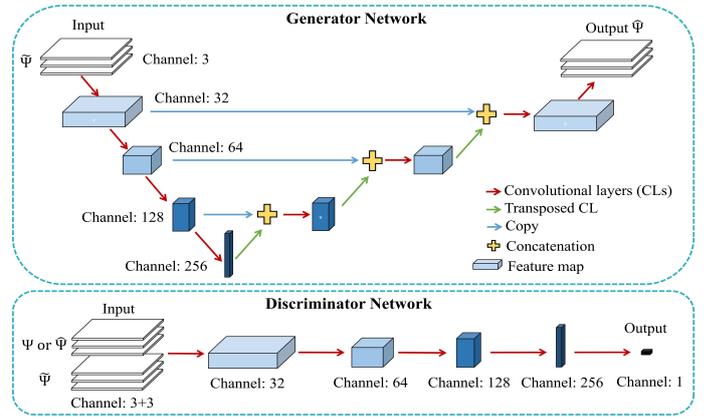}
\caption{The structure of the proposed generative model.}
\label{network}
\end{figure}

It should be noted that, the proposed generative model exploits the advanced structures of U-net, i.e. the symmetric upsampling and downsampling parts with skip connection on the generator. Hence, the area corresponding to $\mathcal{C}_{\mathbf{z}_{\mathrm{O}}}$ occupied by obstacles can be better segmented \cite{levie2021radiounet}. In addition, the skip connection enables different receptive field sizes corresponding to different layers, so that the proposed model can deal with the sensing of obstacles of different sizes. The structure of the proposed generative model is summarized in Fig. \ref{network}.


According to the adversarial training and the advanced structures, to facilitate the solving of problem (\ref{reconstruction_problem}), a corresponding adversarial loss $V_{\mathrm{ad}}(D,G)$ \cite{gulrajani2017improved} should be added to the objective function of (\ref{reconstruction_problem}). That is, we propose a novel training object for our CGAN-based generative model, where the object consists the functions  $V_{\mathrm{MSE}}(G)=\mathbb{E}_{\widehat{\boldsymbol{\Psi}}_{\mathbf{z}_{\mathrm{O}}}}[f(G(\widetilde{\boldsymbol{\Psi}}_{\mathbf{z}_{\mathrm{O}}}))]$ and $V_{\mathrm{ad}}(D,G)$, so as to pay more attention to obstacle areas and minimize the weighted MSE with adversarial training. As a result, the training of the proposed model follows
\begin{equation}
\underset{\mathbf{\Theta}_{g}}{\mathrm{min}}\,\underset{\mathbf{\Theta}_{d}}{\mathrm{max}}\,V(D, G)=V_{\mathrm{ad}}(D,G)+1000 V_{\mathrm{MSE}}(G),
 \label{objective function}
\end{equation}where $V_{\mathrm{ad}}(D,G)=-10\mathbb{E}_{\boldsymbol{\Psi}_{\mathbf{z}_{\mathrm{O}}}'}[(\|\nabla_{\boldsymbol{\Psi}_{\mathbf{z}_{\mathrm{O}}}'} D(\boldsymbol{\Psi}_{\mathbf{z}_{\mathrm{O}}}' \vert \widetilde{\boldsymbol{\Psi}}_{\mathbf{z}_{\mathrm{O}}})\|_{2}-1)^{2}]+\mathbb{E}_{\boldsymbol{\Psi}_{\mathbf{z}_{\mathrm{O}}}}[D(\boldsymbol{\Psi}_{\mathbf{z}_{\mathrm{O}}} \vert \widetilde{\boldsymbol{\Psi}}_{\mathbf{z}_{\mathrm{O}}})]-\mathbb{E}_{\widehat{\boldsymbol{\Psi}}_{\mathbf{z}_{\mathrm{O}}}}[D(\widehat{\boldsymbol{\Psi}}_{\mathbf{z}_{\mathrm{O}}} \vert \widetilde{\boldsymbol{\Psi}}_{\mathbf{z}_{\mathrm{O}}})]$. Note that $\boldsymbol{\Psi}_{\mathbf{z}_{\mathrm{O}}}'=p_{\mathrm{m}}\widehat{\boldsymbol{\Psi}}_{\mathbf{z}_{\mathrm{O}}} +(1-p_{\mathrm{m}})\boldsymbol{\Psi}_{\mathbf{z}_{\mathrm{O}}}$, $\widehat{\boldsymbol{\Psi}}_{\mathbf{z}_{\mathrm{O}}} =G(\widetilde{\boldsymbol{\Psi}}_{\mathbf{z}_{\mathrm{O}}})$, and \(p_{\mathrm{m}} \sim \mathcal{U}(0,1)\). By using the training algorithm, i.e. Algorithm 1 in \cite{gulrajani2017improved}, we can obtain the trained model after a number of training epochs.

\subsection{Online Obstacle Sensing}

For online obstacle sensing, the generator \(G\) of the trained generative model is utilized. Specifically, the THz-based radio environment awareness is first performed by inference on a subset of the radio environment (i.e. the sparse RSS measurements $\widetilde{\boldsymbol{\Psi}}_{\mathbf{z}_{\mathrm{O}}}$) obtained in practice. After compressing the dimensions of the generated matrix from three channels to one channel, the obstacle information can be obtained by the strength of the grid values at different locations on the resulting matrix. That is, we can segment the girds whose value satisfies  $\Psi_{\mathrm{smax}}<\widehat{\Psi}(\mathbf{x}')\le 1$, as the grid $\mathbf{x}'$ is covered by an obstacle based on our preprocessing. In addition, the aware radio environment can be obtained by the inverse operation of preprocessing and the replacement of known RSS values $\{\Psi_{\mathbf{z}_{\mathrm{O}}}\left(\mathbf{x}_{\mathrm{rx}, k}\right)\}_{k=1}^{K}$. Note that if it is used for performance evaluation, the weights for calculating the distortion function can be known.


\section{Performance Evaluation}

The considered 2D scenarios and the corresponding radio environments are both obtained by ray tracing \cite{remcom}. Specifically, by considering THz channel characteristics, each ray is allowed to be reflected and diffracted at most once, as the THz signal suffers from molecular absorption and severe path loss. For the transmitted THz signal, the beamwidth and the carrier frequency are set to $\theta_{\mathrm{b}}=20^{\circ}$ \cite{8880819} and $f_{\mathrm{c}}=300 \mathrm{GHz}$, respectively. In parallel, each scenario is of size $16\mathrm{m}\times20\mathrm{m}$ and divided into $N_{\mathrm{A}}= 48 \times 48=2304$ grids. The training data for the proposed CGAN-based generative model consists of $4500$ deployments of obstacles, where the number of obstacles is randomly selected from $\{1,2,3,4,5\}$. Similarly, for online obstacle sensing when performing the evaluation, the testing data consists of $900$ deployments of $6$ obstacles. The subset of each THz-based radio environment is obtained by randomly sampling from each corresponding complete radio environment, where the sampling rate is set to $0.5$ unless otherwise stated. Note that as the method proposed by \cite{zeng2022uav} required evenly distributed samples, which may be difficult to achieve in practice, we only compare the proposed framework with \cite{hu20233d,levie2021radiounet,ronneberger2015u}.


\begin{figure}[!t]
\center
\includegraphics[width=6.3cm,height=3.2cm] {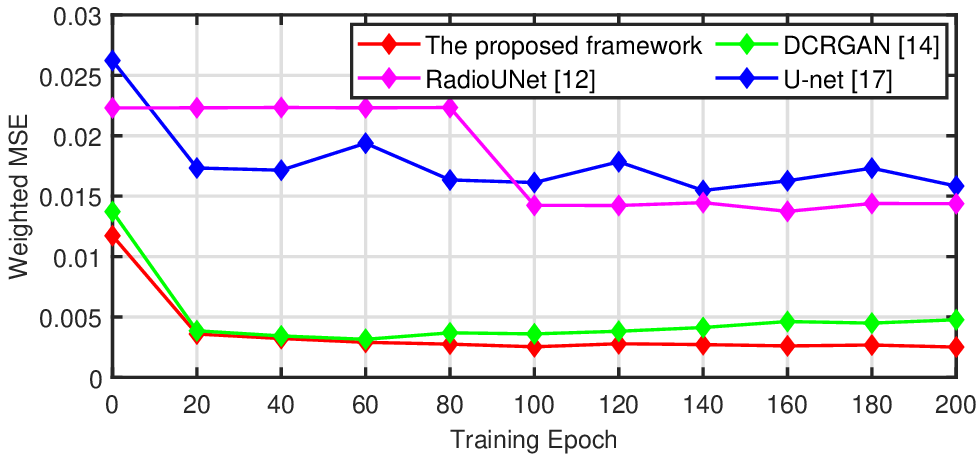}
\caption{Weighted MSE as a function of training epoch.}
\label{passive_MSE}
\end{figure}

\begin{figure}[!t]
\center
\includegraphics[width=6.1cm,height=3.2cm] {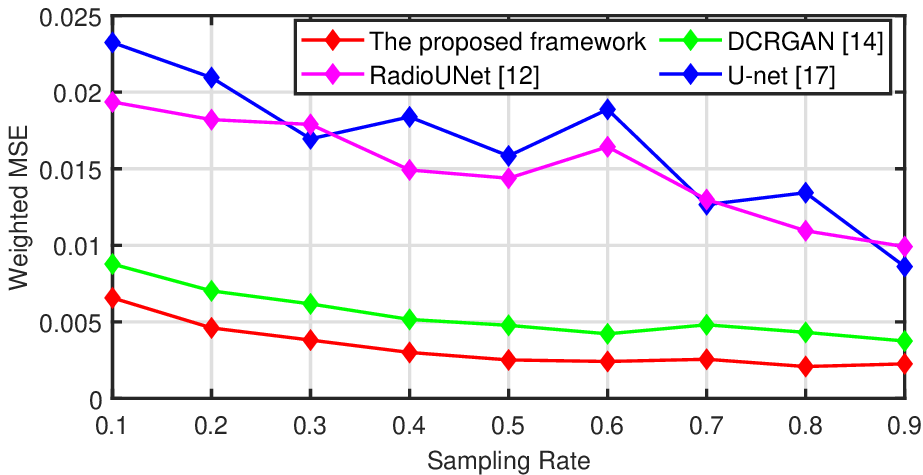}
\caption{Weighted MSE as a function of sampling rate.}
\label{sampling_MSE}
\end{figure}

\subsection{THz-based Radio Environment Awareness}

For the evaluation of THz-based radio environment awareness, Fig. \ref{passive_MSE} investigates the weighted MSE between the complete and the aware radio environments as a function of training epoch. From Fig. \ref{passive_MSE}, we notice that the MSE curve of DCRGAN presents an overfitting phenomenon, which may result from its failure to adapt to little a priori knowledge. Moreover, it can be observed that the weighted MSE corresponding to the baselines are all higher than that achieved by the proposed framework, which indicates the superior awareness performance of the proposed framework and can result in better obstacle sensing performance due to the segmentation.

Considering the impact of the amount of measured RSS data in each subset of the THz-based radio environment, Fig. \ref{sampling_MSE} investigates the weighted MSE as a function of sampling rate. It can be observed that as the sampling rate increases, the performance of radio environment awareness is improving. An essential reason is that the measured RSS can reflect more obstacle information with the highly directional THz beams. Fig. \ref{sampling_MSE} reveals that, even if the availability of the sparsely RSS data is poor, the proposed framework can obtain more accurate awareness of the radio environment than baselines.

\subsection{THz-based Obstacle Sensing}

\begin{figure}[!t]
\center
\includegraphics[width=6.5cm,height=3.2cm] {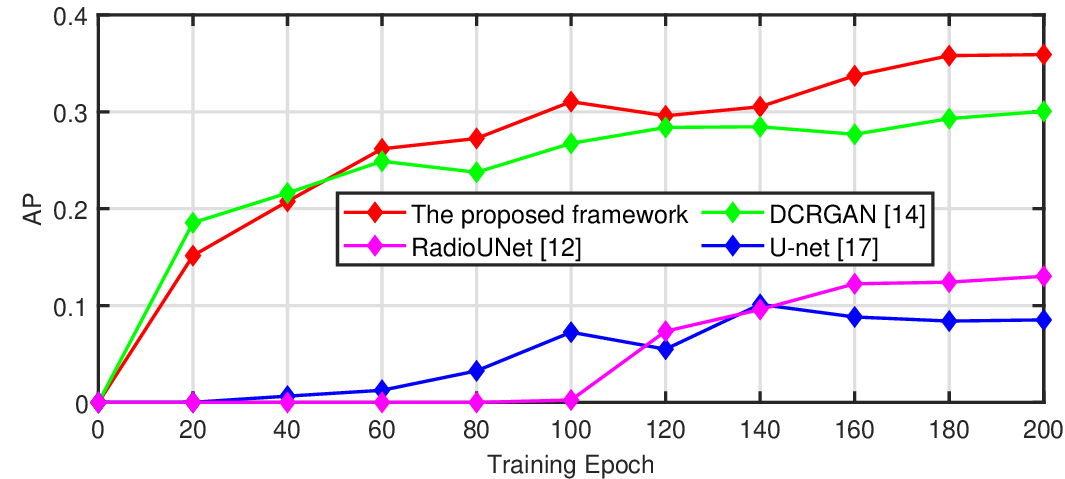}
\caption{AP as a function of training epoch.}
\label{passive_AP}
\end{figure}

\begin{figure}[!t]
\center
\includegraphics[width=6.2cm,height=3.2cm] {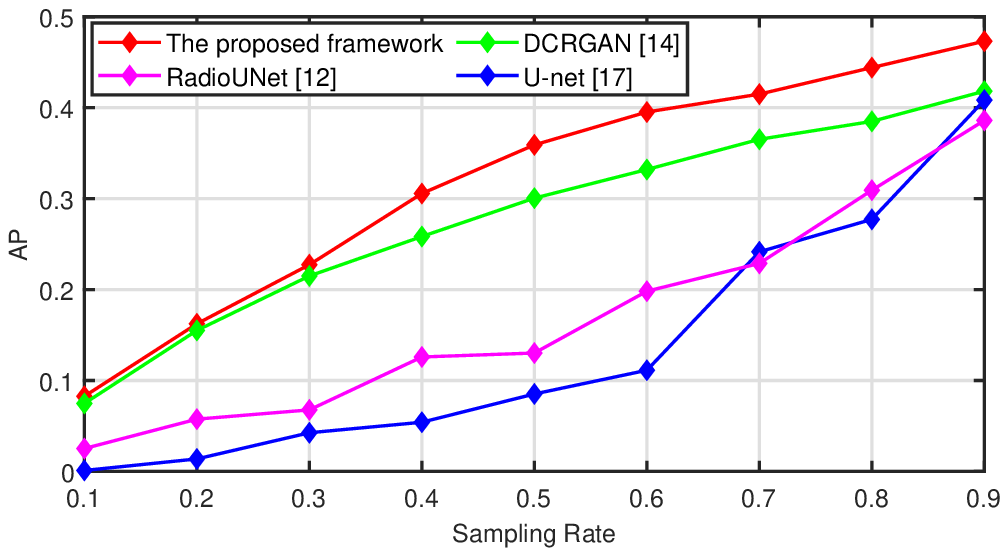}
\caption{AP as a function of sampling rate.}
\label{sampling_AP}
\end{figure}

For the evaluation of THz-based obstacle sensing, Fig. \ref{passive_AP} and Fig. \ref{sampling_AP} investigate the AP as a function of training epoch and sampling rate, respectively. Specifically, AP is a commonly used metric in the field of object detection, which measures the overlap between the detected target range and the true target range from both recall and precision perspectives. Therefore, we borrow this metric to measure the obstacle sensing performance, where a higher AP value indicates a greater overlap, i.e. the better sensing of obstacles' location and shape. Therefore, observations can be given in Fig. \ref{passive_AP} and Fig. \ref{sampling_AP} that the proposed framework outperforms baselines in terms of AP regarding obstacles' shape and location, and the performance of obstacle sensing is also improved with the increase in sampling rate.

\begin{figure}[!t]
\center
\includegraphics[width=8.5cm,height=10.5cm] {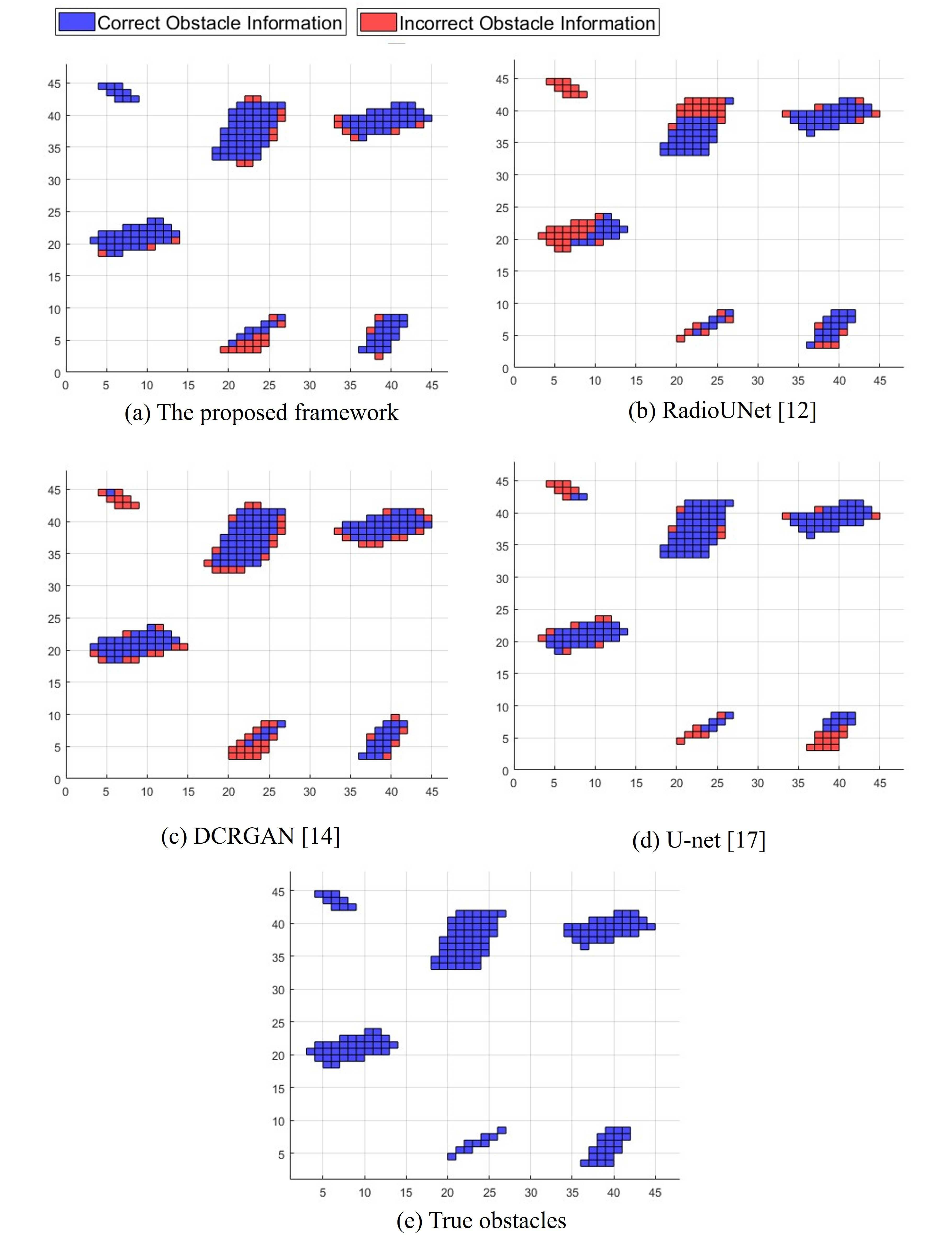}
\caption{An example of the true obstacles and the sensed obstacles obtained by the proposed framework and baselines, where the sensing errors are indicated by red in the sensed obstacle.}
\label{new_obstacle}
\end{figure}


In Fig. \ref{new_obstacle}, we give an example to compare the true and the sensed obstacles through the aware THz-based radio environment, where the sensing errors are indicated by red in the sensed obstacle. It can be seen from Fig. \ref{new_obstacle} that only the proposed framework can sense the obstacle in the upper left corner without error, and our proposed framework can characterize the deployment and shape information of obstacles very well due to a small number of red grids, which is lower than that presented by other baselines. Note that since it's just an example, it can be inconsistent with the aforementioned averaged results.

\section{Conclusions}

In this paper, we have studied a THz-based radio environment awareness problem for obstacle sensing, where the expectation of awareness performance corresponding to the probability distribution of obstacle attributes has been considered. To solve this problem, a generative radio environment awareness framework and the corresponding generative model have been proposed. Simulation results show that the proposed scheme can effectively perform THz-based radio environment awareness and obstacle sensing with the lowest MSE and the highest AP compared to other methods.

\bibliographystyle{IEEEtran}
\bibliography{IEEEabrv,ref}

\end{document}